\documentclass[letterpaper]{article} % DO NOT CHANGE THIS
    \usepackage[preprint]{aaai2027}  % DO NOT CHANGE THIS
    \usepackage[hyphens]{url}  % DO NOT CHANGE THIS
    \usepackage{graphicx} % DO NOT CHANGE THIS
    \usepackage{natbib}  % DO NOT CHANGE THIS AND DO NOT ADD ANY OPTIONS TO IT
    \usepackage{caption} % DO NOT CHANGE THIS AND DO NOT ADD ANY OPTIONS TO IT
    \usepackage{subcaption}
    \usepackage{algorithm}
    \usepackage{algorithmic}
    \usepackage{amsmath}
    \usepackage{amssymb}
    \usepackage{booktabs}
    \usepackage{multirow}
    \usepackage{graphicx}
    
    \newcommand{\optionalgraphic}[2][]{%
        \IfFileExists{#2}{\includegraphics[#1]{#2}}{%
            \fbox{\parbox[c][0.16\textheight][c]{0.9\columnwidth}{%
                \centering Missing figure\\[2pt]\texttt{\detokenize{#2}}}}}}
    \usepackage{booktabs}
    
    \title{ProtoGIB-Workload: Learning Workload-Specific Neural Topology Prototypes across Subjects}
    \author{
Yuzhe Zhang\textsuperscript{\rm 1}\equalcontrib,
Yixi Zhang\textsuperscript{\rm 1}\equalcontrib,
Shengdian Jiang\textsuperscript{\rm 1}\equalcontrib,
Chengxi Xie\textsuperscript{\rm 2},
Jihong Wang\textsuperscript{\rm 2}\corresponding,
Huan Liu\textsuperscript{\rm 2},
Man Yao\textsuperscript{\rm 3},
Minnan Luo\textsuperscript{\rm 2},
Chao Shen\textsuperscript{\rm 2}
}

\affiliations{
\textsuperscript{\rm 1}College of Artificial Intelligence,
Nanjing University of Aeronautics and Astronautics\\
\textsuperscript{\rm 2}School of Computer Science and Technology,
Xi'an Jiaotong University\\
\textsuperscript{\rm 3}Institute of Automation,
Chinese Academy of Sciences
}
    
\begin{document}
    
    \maketitle
    
    \begin{abstract}
Reliable electroencephalography (EEG)-based mental workload recognition is crucial for adaptive human-centered systems, yet practical deployment requires models to generalize to users unseen during training. Although functional connectivity graphs are widely adopted to capture workload-related neural interactions, they inherently entangle task-relevant structures with subject-specific physiological traits and sample-level noise. This entanglement often leads models to learn structural shortcuts, severely degrading cross-subject generalization. To address this, we propose \textbf{ProtoGIB-Workload}, a novel framework that explicitly regularizes and aligns graph structures for subject-independent workload recognition. Our approach introduces a Stochastic Graph Information Bottleneck (SGIB) to compress dense correlation priors into compact, task-relevant subgraphs, filtering out input-related redundancy. Crucially, to prevent the retention of subject-specific spurious edges, we propose a Class-Conditional Topology Stabilizer (CTS). Leveraging the fixed electrode coordinates of EEG data, CTS operates directly on graph-generation probabilities to encourage consistent edge-generation statistics across different subjects sharing the same workload class. Extensive experiments on two public EEG workload datasets and one in-house EEG cognitive load dataset of air traffic controllers under strict leave-one-subject-out (LOSO) protocols demonstrate that ProtoGIB-Workload significantly outperforms state-of-the-art temporal and graph-based baselines, improving the cross-subject Macro-F1 score by an average of 5.15\% (up to 6.34\%). Further analyses confirm that our method successfully extracts stable, cross-subject consistent neural connectivity patterns.
    \end{abstract}
    
    % Uncomment the following to link to your code, datasets, an extended version or similar.
    % You must keep this block between (not within) the abstract and the main body of the paper.
    % Make sure that you do not de-anonymize yourself with these links.
    % \begin{links}
    %     \link{Code}{https://aaai.org/example/code}
    %     \link{Datasets}{https://aaai.org/example/datasets}
    %     \link{Extended version}{https://aaai.org/example/extended-version}
    % \end{links}
    \section{Introduction}
Mental workload profoundly influences human performance in safety-critical systems~\cite{young2015mental_workload}. While electroencephalography (EEG) provides direct sensitivity to neural dynamics for continuous monitoring~\cite{borghini2014neurophysiological}, its scalable deployment is severely hindered by the need for tedious individual calibration. Therefore, achieving \emph{subject-independent generalization} to operate robustly on unseen users without prior data collection is imperative~\cite{wu2022transfer_eeg}. However, this remains a formidable challenge: inter-subject variability, driven by differences in neuroanatomy and physiological traits, leads to highly heterogeneous EEG distributions even under identical workload conditions.

To characterize task-relevant neural dynamics amidst this variability, functional connectivity graphs are widely used to capture inter-regional brain coordination~\cite{safari2024workload_connectivity,wei2025sgatcns}. However, directly relying on raw functional graphs introduces a critical challenge: observed EEG graphs entangle task-relevant structures with stable subject-specific physiological traits. Neuroscientific studies show functional connectivity acts as an ``individual signature,'' where connectivity from the same subject across different tasks is often more similar than from different subjects performing the identical task~\cite{finn2015connectome,gratton2018stable_networks,nentwich2020eeg_fc}. Consequently, models may exploit these traits as structural shortcuts, degrading cross-subject generalization.

\begin{figure}[!t]
\centering
\includegraphics[width=1.0\columnwidth]{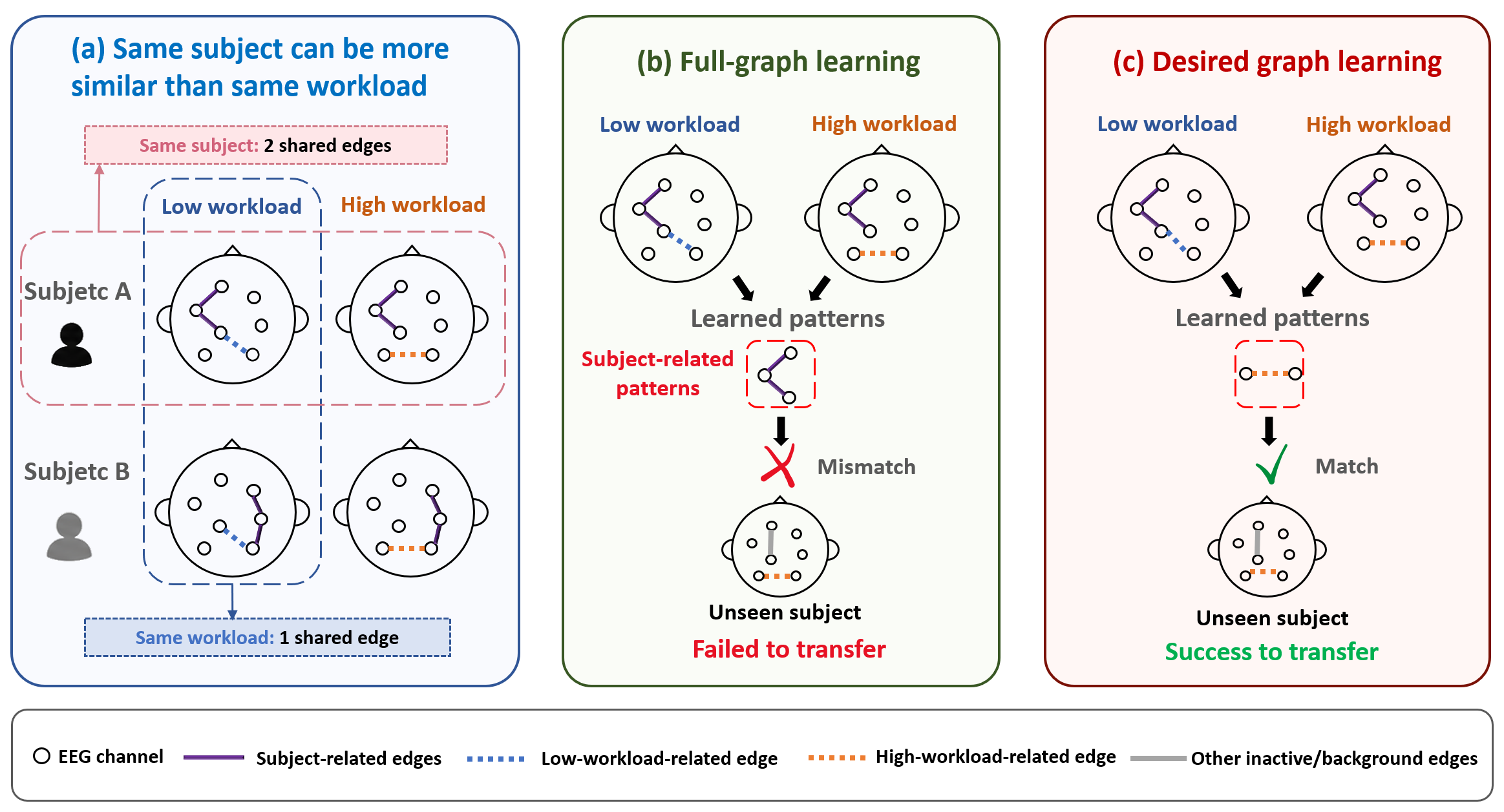} % Reduce the figure size so that it is slightly narrower than the column. Don't use precise values for figure width.This setup will avoid overfull boxes.
\caption{Subject-specific connectivity and cross-subject stable patterns in EEG workload recognition.}
\label{fig:motivation}
\end{figure}

To mitigate this, introducing Information Bottleneck (IB) theory to graph learning (e.g., stochastic graph bottlenecks~\cite{wu2020gib,miao2022gsat,yan2025federated}) is a principled approach. By constraining mutual information between input features and generated subgraphs, they discard structural redundancies at the single-sample level. However, overcoming cross-subject heterogeneity requires addressing two distinct statistical hierarchies: individual structural compression and population-level topology consistency. While single-sample compression filters out isolated physiological noise, it does not ensure that the resulting topologies for the same workload state align across the population (Fig. \ref{fig:motivation}). Because subjects possess inherently diverse baseline connectomes, individually compressed graphs often suffer from class-conditional topology shifts. This motivates us to explicitly regularize the graph-generation process from a population perspective. Fortunately, because EEG recordings across subjects share a fixed electrode coordinate system, it provides a natural structural constraint position to directly compare and stabilize edge-generation statistics at corresponding anatomical locations~\cite{seeck2017standardized}.

To address this problem, we propose \textbf{ProtoGIB-Workload}, a unified graph bottleneck framework for subject-independent EEG workload recognition. It achieves robust generalization through two synergistic mechanisms operating at distinct statistical levels. First, at the micro-level, a Stochastic Graph Information Bottleneck (SGIB) applies the graph information bottleneck principle~\cite{fu2025discrete} to dynamically compress dense functional-correlation priors into compact, task-relevant subgraphs for each sample, filtering out redundant physiological connectivity. Second, at the macro-level, we propose the Class-Conditional Topology Stabilizer (CTS) to ensure cross-subject topological consistency. By exploiting the spatial correspondence of EEG sensors, CTS aggregates soft edge-retention probabilities into subject-class mean profiles and contracts them toward workload-specific topology prototypes under subject-balanced weighting. A separation objective further prevents different workload classes from collapsing into an indistinguishable topology. We show that this population contraction controls an upper bound on the source-subject information represented by a class-conditional mean-field topology surrogate.

We evaluate ProtoGIB-Workload on two public EEG datasets and one in-house EEG cognitive load dataset of air traffic controllers under strict leave-one-subject-out (LOSO) protocols. Extensive experiments demonstrate that our approach consistently outperforms representative baselines, achieving an average improvement of 5.15\% in cross-subject Macro-F1 score (with a notable 6.34\% gain on EEGMAT). Connectivity analyses confirm the learned prototypes exhibit structural consistency across subjects within the same class. The main contributions are:

\begin{itemize}
\item \textbf{Dual-Level Topological Perspective}: We formulate subject-independent EEG workload recognition by distinguishing two statistical hierarchies: single-sample structural compression and population-level topology stabilization, providing a comprehensive perspective to tackle subject-specific topological shifts.

\item \textbf{Unified Graph Bottleneck Framework}: We propose ProtoGIB-Workload, integrating a Stochastic Graph Information Bottleneck (SGIB) to extract compact subgraphs individually, and a Class-Conditional Topology Stabilizer (CTS) that exploits the common electrode coordinate system to contract subject-class mean edge-retention profiles toward stable, class-specific topology prototypes.

\item \textbf{Principled Analysis and Evaluation}: We establish a variational bound connecting prototype contraction to the source-subject information represented by class-conditional mean-field topology statistics, and validate the framework under strict unseen-subject protocols.
\end{itemize}

    \section{Related Work}
\label{sec:related-work}

\noindent\textbf{EEG-based Workload Recognition.}
EEG workload recognition has been traditionally approached from temporal and graph perspectives. Temporal models focus on sequence dynamics, evolving from compact CNNs like EEGNet~\cite{lawhern2018eegnet} to advanced Transformer architectures~\cite{song2023eegconformer,yan2025biconformer,ding2025eegdeformer,wang2024arfn}. Meanwhile, graph neural networks (GNNs) are widely adopted to capture cross-region functional connectivity. For instance, some approaches learned sample-dependent adjacency~\cite{song2020dgcnn,song2020instanceadaptive}, incorporated domain-adversarial training~\cite{zhong2022rgnn}, fused spatial priors with adaptive attention~\cite{philipchen2025adamgraph}, or estimated directed topology with a sparse adjacency~\cite{xiao2025dcgnn,liu2024vbhgnn}. Despite these advances, under a strict leave-one-subject-out (LOSO) protocol, these methods do not explicitly enforce class-conditional cross-subject consistency during graph generation, leaving the model vulnerable to learning subject-specific structural shortcuts.

\noindent\textbf{Graph Information Bottleneck and Topology Regularization.}
Graph information bottleneck (GIB) methods~\cite{wu2020gib,miao2022gsat,yan2025federated} extract compact subgraphs by stochastically restricting information flow. In parallel, invariant-subgraph approaches study structural robustness under generic distribution shifts~\cite{liu2025subgraph,gong2026invariant}, and topology-aware generation aims to preserve global graph characteristics during synthesis~\cite{ParJoo_Conditional_MICCAI2025}. While these methods effectively address \emph{sample-level} structural compression and general OOD generation, individually optimal graph generation does not guarantee topology alignment across a population with diverse baseline connectomes. To explicitly address severe inter-subject variability in EEG, our framework introduces CTS to model population-level topology. Operating entirely on differentiable graph-generation probabilities, CTS constructs subject-class population statistics and stabilizes them through a dual-objective design: a subject-balanced contraction pulling subject-class mean edge-retention profiles toward non-trainable exponential moving average (EMA) prototypes, and a separation objective preventing class-level topology collapse. This ensures cross-subject structural consistency at the probabilistic generation stage.

    \section{Method}
    
\subsection{Framework Overview}

\begin{figure*}[t]
\centering
\includegraphics[width=0.9\textwidth]{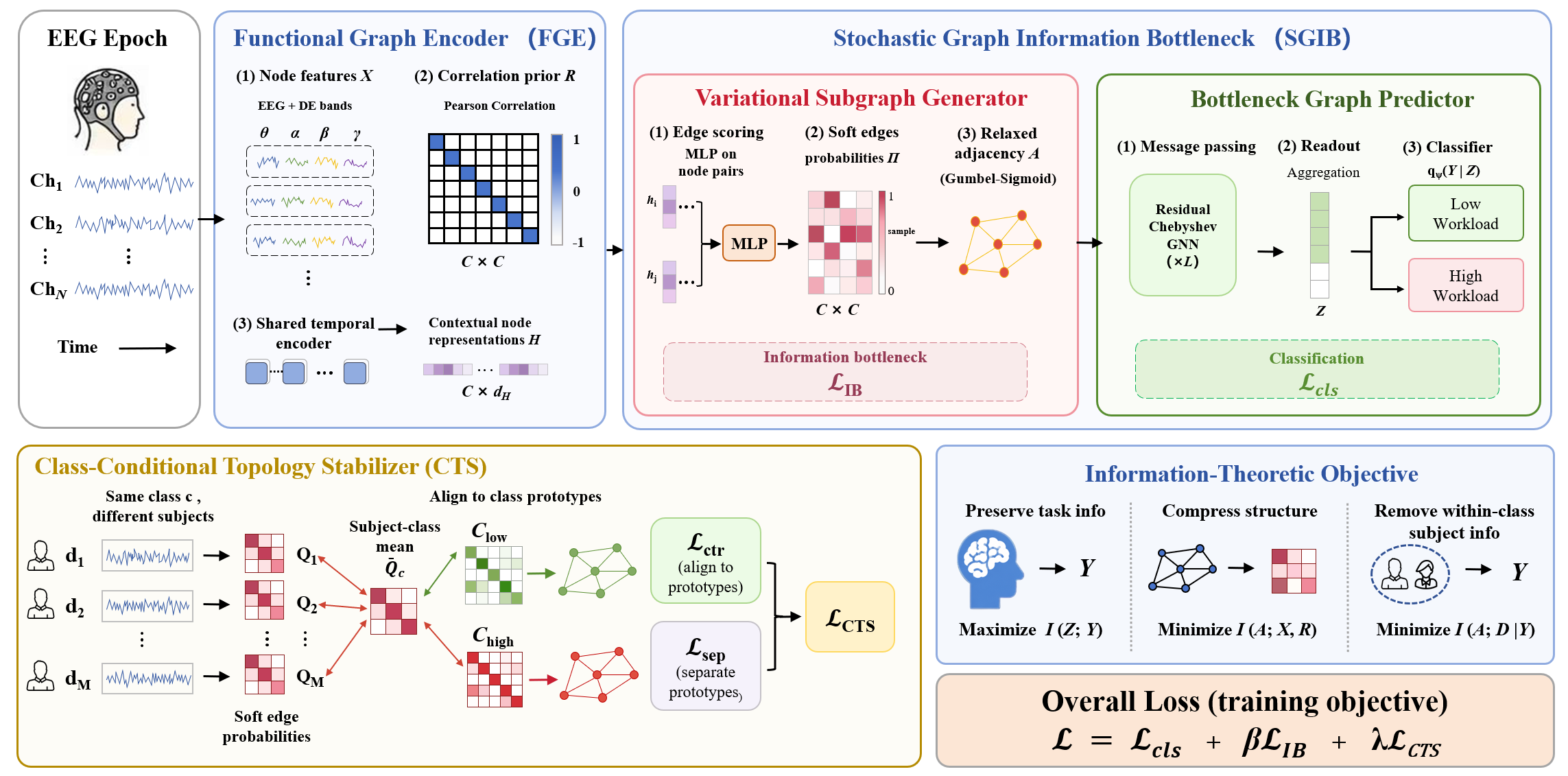}
\caption{The overall architecture of the proposed ProtoGIB-Workload framework. The framework consists of a Stochastic Graph Information Bottleneck (SGIB) for extracting compact, task-relevant subgraphs from dense functional priors, and a Class-Conditional Topology Stabilizer (CTS) that aligns subject-level edge-retention profiles toward workload-specific prototypes to ensure cross-subject stability.}
\label{fig:framework}
\end{figure*}

Given an EEG epoch \(E\in\mathbb{R}^{C\times T}\), let \(Y\in\{1,\ldots,K\}\) be its corresponding mental workload label and \(D\in\{1,\ldots,N\}\) denote the subject identity. We use the fixed undirected edge universe \(\mathcal{E}_0=\{\{u,v\}:1\leq u<v\leq C\}\) for every sample; all latent adjacency matrices are generated within this common coordinate system.  As illustrated in Figure~\ref{fig:framework}, the proposed ProtoGIB-Workload framework comprises three main modules: a Functional Graph Encoder (FGE), a Stochastic Graph Information Bottleneck (SGIB), and a Class-Conditional Topology Stabilizer (CTS). First, the FGE computes a dense correlation prior from the raw EEG signals and extracts channel-wise temporal-spectral representations. Second, the SGIB dynamically infers a compact, task-relevant subgraph and performs workload prediction. The SGIB itself consists of two internal components: a Variational Subgraph Generator (VSG) that samples the latent adjacency matrix, and a Bottleneck Graph Predictor (BGP) that performs message passing over the generated topology. Finally, the CTS acts as a structural regularizer during training, encouraging source subjects sharing the same workload label to generate consistent underlying structures.

Let \(X\) represent the node features, \(R\) the dense statistical prior, \(A\) the discrete latent adjacency generated by SGIB, and \(Z\) the aggregated graph-level representation. In our formulation, the Variational Subgraph Generator (VSG) within SGIB serves as the generative distribution \(q_{\phi}(A\mid X, R)\), while the Bottleneck Graph Predictor (BGP) acts as the predictive distribution \(q_{\psi}(Y\mid Z)\). Guided by the principles of variational and graph information bottlenecks~\cite{alemi2017deep,wu2020gib}, we cast the objective of generating cross-subject stable subgraphs into three complementary information-theoretic terms:
\begin{align}
\max_{\phi,\psi}\;\mathcal{J}
&=I(Z;Y)-\beta I(A;X,R)
-\lambda I(A;D\mid Y).
\label{eq:info_objective}
\end{align}

The first term, \(I(Z;Y)\), preserves predictive information for workload discrimination. The second, \(I(A;X,R)\), compresses structure by penalizing feature-dependent dense graphs, forming an information bottleneck between node features (and prior) and topology. The third, \(I(A;D\mid Y)\), measures class-conditional subject information in topology; minimizing it stabilizes mean edge-retention statistics within each workload class. The following sections introduce tractable surrogates for these terms.

\subsection{Functional Graph Encoder}

The FGE module transforms the raw EEG epoch \(E\) into node representations \(X\). Each channel's input \(x_u\) concatenates the raw time-series signal with differential-entropy descriptors from the \(\theta, \alpha, \beta, \gamma\) frequency bands~\cite{zheng2015seed}. A shared temporal encoder \(f_{\mathrm{temp}}\) then extracts localized temporal-spectral patterns:
\begin{equation}
h_u=f_{\mathrm{temp}}(x_u),
\qquad u=1,\ldots,C.
\label{eq:node_encoding}
\end{equation}
Simultaneously, to incorporate neurophysiological priors, we compute the channel-wise Pearson correlation coefficient (PCC) matrix of \(E\). Unlike prior works using this as a rigid topology~\cite{safari2024workload_connectivity}, we treat it as a dense statistical prior \(R\), where \(R_{uv} = |\operatorname{PCC}(E)_{uv}|\). The node embeddings \(H = \{h_1, \dots, h_C\}\) and the prior \(R\) are subsequently fed into the downstream information bottleneck to dynamically infer task-relevant structures.

\subsection{Stochastic Graph Information Bottleneck}

Functional brain networks constructed from EEG signals inherently contain dense, spurious connections and physiological noise irrelevant to cognitive workload. Directly applying Graph Neural Networks (GNNs) over such dense topologies often leads to overfitting and suboptimal generalization. To address this, we introduce a Stochastic Graph Information Bottleneck (SGIB) that explicitly compresses the dense prior into a compact, task-relevant subgraph.

To achieve this structural compression, the Variational Subgraph Generator (VSG) first evaluates every unordered edge \(\{u,v\}\in\mathcal{E}_0\) in the latent space. To fuse the learned physiological features with the statistical prior while ensuring edge symmetry, we model the edge probability using a symmetrized Multilayer Perceptron (MLP):
\begin{equation}
\begin{aligned}
\phi_{uv} &= \operatorname{MLP}_{\phi}([h_u \parallel h_v \parallel R_{uv}]), \\
\pi_{uv} &= \sigma\!\left( \frac{\phi_{uv} + \phi_{vu}}{2} \right),
\end{aligned}
\label{eq:edge_probability}
\end{equation}
where \(\parallel\) denotes concatenation. The symmetric matrix \(\Pi=[\pi_{uv}] \in [0,1]^{C \times C}\) yields the vectorized edge probabilities \(Q = \operatorname{vec}_{\mathcal{E}_0}(\Pi) \in [0,1]^{|\mathcal{E}_0|}\). We assume the generative distribution factorizes independently over edges: \(q_{\phi}(A\mid X,R) = \prod_{\{u,v\}\in\mathcal{E}_0} \operatorname{Bern}(a_{uv}; \pi_{uv})\).

To enable backpropagation through the discrete binary decision \(a_{uv}\sim\operatorname{Bern}(\pi_{uv})\), we employ the standard continuous Gumbel-Sigmoid relaxation and Straight-Through (ST) estimator~\cite{maddison2017concrete} (details in supplementary). This yields a symmetric discrete adjacency matrix \(A_{\text{ST}}\) during the forward pass while allowing continuous gradients during the backward pass.

To enforce the structural compression objective \(-I(A;X,R)\), the VSG must discard redundant connections. By introducing an input-independent sparse prior \(r(A)=\prod_{e\in\mathcal E_0}\operatorname{Bern}(A_e;\rho)\) (\(\rho \ll 1\)), we upper-bound the mutual information via the Kullback-Leibler (KL) divergence~\cite{miao2022gsat}:
\begin{equation}
\mathcal{L}_{\mathrm{IB}}
=\frac{1}{|\mathcal{E}_0|}
\mathbb{E}_{X,R}
\sum_{e\in\mathcal{E}_0}
\operatorname{KL}\!\left(
\operatorname{Bern}(\pi_{e})
\Vert
\operatorname{Bern}(\rho)
\right).
\label{eq:ib_loss}
\end{equation}
This variational penalty regularizes topology by pruning input-dependent edges into a compact bottleneck graph.

Given the stochastically sampled \(A_{\text{ST}}\), the BGP translates the sparse topology into the workload prediction. We employ residual Chebyshev graph convolutions~\cite{defferrard2016chebnet} as the \(\operatorname{GNN}\) backbone to update node embeddings \(H^g = \operatorname{GNN}(H, A_{\text{ST}})\). A global \(\operatorname{READOUT}\) function then aggregates \(H^g\) into a unified graph-level representation \(Z\), which is mapped to the workload prediction \(\hat{Y}\) via a classifier.

The classifier defines the predictive distribution \(q_{\psi}(Y\mid Z)\) and is optimized via the cross-entropy loss, \(\mathcal{L}_{\mathrm{cls}}=-\mathbb{E}\log q_{\psi}(Y\mid Z)\). Minimizing this cross-entropy maximizes a variational lower bound on \(I(Z;Y)\), encouraging the resulting graph representation to retain predictive information for workload discrimination.

\subsection{Class-Conditional Topology Stabilizer}

While the SGIB successfully extracts compact and task-relevant subgraphs, structural compression alone does not inherently mitigate subject biases. Even compact subgraphs may exhibit systematic subject-dependent variations. The Class-Conditional Topology Stabilizer (CTS) therefore acts as the second pillar of our framework, encouraging cross-subject consistency among the generated latent topologies sharing the same workload label. Because EEG electrodes maintain clear physical correspondence across subjects, we can directly regularize the edge-generation statistics at these fixed anatomical locations. Its information-theoretic target is the conditional mutual information:
\begin{equation}
I(A;D\mid Y)
=\mathbb{E}_{y,d}
\operatorname{KL}\!\left(
p(A\mid y,d)\Vert p(A\mid y)
\right).
\label{eq:conditional_mask_mi}
\end{equation}
This quantity measures the discrepancy between the subject-conditioned latent edge distribution \(p(A\mid y,d)\) and the class-level mixture \(p(A\mid y)\). Conditioning on \(Y\) targets within-class source-subject variation without directly aligning topologies from different workload classes.

Directly optimizing Eq.~\eqref{eq:conditional_mask_mi} is intractable due to the exponentially large discrete graph space. To make the analysis tractable, we adopt a mean-field assumption~\cite{dan2026vmf,miao2022gsat}, approximating the joint distribution using independent edge probabilities. We establish the following bound (proof in supplementary):

\noindent\textbf{Theorem 1 (Conditional Mean-Field Subject-Information Bound).}
\textit{Let \(\mu_{c,d}=\mathbb{E}[Q\mid Y=c,D=d]\) be the first-order edge statistics, and \(\bar I_{\mathrm{MF}}(A;D\mid Y=c)\) be the corresponding class-conditional mean-field subject information. Let \(C_c\in[\delta,1-\delta]^{|\mathcal E_0|}\) be any class reference topology for some \(\delta\in(0,1/2)\). Then}
\begin{equation}
\bar I_{\mathrm{MF}}(A;D\mid Y=c)
\leq
\frac{1}{\delta(1-\delta)}
\mathbb{E}
\!\left[\|Q-C_c\|_2^2\mid Y=c\right].
\label{eq:conditional_mi_bound}
\end{equation}

\begin{table*}[!t]
\centering
\footnotesize
\setlength{\tabcolsep}{2.5pt}
\renewcommand{\arraystretch}{0.9}

\resizebox{\textwidth}{!}{
\begin{tabular}{llcccccc}
\toprule

\multirow[c]{2}{*}{Category}
&
\multirow[c]{2}{*}{Method}
&
\multicolumn{2}{c}{STEW}
&
\multicolumn{2}{c}{EEGMAT}
&
\multicolumn{2}{c}{SELF}
\\

\cmidrule(lr){3-4}
\cmidrule(lr){5-6}
\cmidrule(lr){7-8}

& & ACC & F1 & ACC & F1 & ACC & F1 \\

\midrule

\multirow[c]{4}{*}{General EEG}
& EEGLearn
& $88.19\pm17.58$ & $85.71\pm22.63$
& $81.94\pm25.42$ & $77.59\pm31.08$
& $55.00\pm16.72$ & $52.87\pm19.62$
\\

& EEGNet
& $80.73\pm19.28$ & $79.21\pm21.57$
& $64.35\pm22.70$ & $63.44\pm22.94$
& $48.61\pm17.03$ & $46.83\pm17.49$
\\

& EEG-Conformer
& $86.63\pm11.69$ & $86.14\pm13.02$
& $71.99\pm17.25$ & $70.99\pm18.17$
& $63.61\pm18.88$ & $61.40\pm20.29$
\\

& EEG-Deformer
& $92.88\pm9.73$ & $92.49\pm11.04$
& $67.36\pm20.26$ & $65.58\pm21.25$
& $61.67\pm17.98$ & $61.71\pm17.52$
\\

\midrule

\multirow[c]{4}{*}{Workload-Specific}
& LSCCN
& $85.33\pm13.39$ & $84.78\pm14.43$
& $65.05\pm19.72$ & $63.92\pm20.47$
& $70.97\pm8.21$ & $69.71\pm7.83$
\\

& MuLHiTA
& $92.45\pm8.58$ & $92.30\pm8.95$
& $80.09\pm15.99$ & $79.56\pm16.44$
& $67.08\pm11.86$ & $66.65\pm12.06$
\\

& EEGMeNet
& \underline{$94.62\pm7.41$}
& \underline{$94.54\pm7.61$}
& \underline{$84.49\pm14.18$}
& \underline{$84.07\pm14.58$}
& $71.81\pm18.12$ & $70.85\pm18.83$
\\

& BiConformer
& $92.27\pm8.59$ & $92.16\pm8.74$
& $67.13\pm21.96$ & $64.98\pm24.14$
& $66.81\pm15.04$ & $66.38\pm15.15$
\\

\midrule

\multirow[c]{4}{*}{Graph-Based}
& DGCNN
& $93.66\pm10.17$ & $93.24\pm11.89$
& $83.33\pm10.76$ & $82.64\pm11.74$
& \underline{$78.75\pm10.30$}
& \underline{$77.91\pm11.23$}
\\

& RGNN
& $91.67\pm11.25$ & $91.05\pm13.30$
& $68.98\pm20.66$ & $67.55\pm21.92$
& $47.50\pm23.07$ & $45.23\pm24.62$
\\

& AdamGraph
& $92.27\pm11.85$ & $92.18\pm12.03$
& $75.93\pm15.69$ & $75.30\pm16.24$
& $65.28\pm11.86$ & $64.11\pm12.16$
\\

& DCGNN
& $87.41\pm14.67$ & $86.90\pm15.89$
& $72.92\pm18.25$ & $72.23\pm18.79$
& $51.25\pm12.55$ & $50.24\pm12.78$
\\

\midrule

Ours
& \textbf{ProtoGIB}
& \boldmath{$96.09\pm6.04$}
& \boldmath{$96.04\pm6.19$}
& \boldmath{$87.50\pm11.20$}
& \boldmath{$87.05\pm11.83$}
& \boldmath{$81.94\pm10.85$}
& \boldmath{$81.62\pm11.31$}
\\

\bottomrule
\end{tabular}
}

\caption{Within-subject comparison on the three workload datasets.}
\label{tab:comparison-ws}
\end{table*}

Theorem 1 provides the theoretical foundation for CTS: contracting subject-level mean topologies toward a shared class reference minimizes an upper bound on the source-subject information retained by the mean-field surrogate. While this framework provides a principled mechanism to suppress subject-dependent variation in first-order edge statistics, we note that it does not claim complete elimination of subject-specific information, particularly concerning higher-order topological dependencies or node-level representations. This design choice represents a calculated trade-off, balancing tractable optimization against the theoretical ideal of strict subject invariance. 

To optimize this bound, CTS regularizes the continuous edge-probability vectors \(Q_i\). We adopt a class-subject-balanced episodic sampler. For a mini-batch, let \(\mathcal{B}_{c,d}\) denote the set of samples belonging to subject \(d\) with workload class \(c\). We compute the differentiable subject-class mean \(\tilde{Q}_{c,d} = \frac{1}{|\mathcal{B}_{c,d}|} \sum_{i \in \mathcal{B}_{c,d}} Q_i\). To provide a stable optimization target \(C_c\), CTS maintains a non-trainable class prototype updated via an exponential moving average (EMA) of the batch-level class means (with clipping to satisfy the \(\delta\) bound in Theorem 1, details in supplementary).

We then use a subject-balanced contraction loss to align sample statistics with the stable EMA prototype:
\begin{equation}
\mathcal{L}_{\mathrm{ctr}}
=\frac{1}{K}
\sum_{c=1}^{K}
\frac{1}{|\mathcal{D}_c|}
\sum_{d\in\mathcal{D}_c}
\|\tilde{Q}_{c,d}-\operatorname{sg}(C_c)\|_2^2,
\label{eq:prototype_contraction}
\end{equation}
where \(K\) is the number of classes, \(\mathcal{D}_c\) is the set of subjects representing class \(c\) in the batch, and \(\operatorname{sg}(\cdot)\) denotes the stop-gradient operation.

Contraction alone may allow the class prototypes to approach the same trivial topology. CTS therefore introduces a separation loss to enforce a minimum structural margin \(m>0\) between the differentiable class means:
\begin{equation}
\mathcal{L}_{\mathrm{sep}}
=\frac{1}{K(K-1)}
\sum_{c\neq c'}
\left[m-\|\tilde{Q}_c-\tilde{Q}_{c'}\|_2\right]_+^2,
\label{eq:sep_loss}
\end{equation}
where \([x]_+=\max(x,0)\). The overall CTS objective couples within-class source-subject consistency with between-class separability as \(\mathcal{L}_{\mathrm{CTS}} = \mathcal{L}_{\mathrm{ctr}} + \gamma\mathcal{L}_{\mathrm{sep}}\).

\subsection{Joint Optimization}

The objective combines task sufficiency, structural compression, and cross-subject topology stabilization:

\begin{equation}
\mathcal{L}
=\mathcal{L}_{\mathrm{cls}}
+\beta\mathcal{L}_{\mathrm{IB}}
+\lambda\mathcal{L}_{\mathrm{CTS}}.
\label{eq:final_loss}
\end{equation}
It is imperative to emphasize that the CTS module, subject identities, and class prototypes are utilized exclusively during training for regularization. During inference, they are completely discarded; the model relies solely on the SGIB to dynamically generate the graph and predict the workload.

    % \input{sections/theoretical-analysis}
    % Requires \usepackage{placeins} in main preamble for \FloatBarrier
\section{Experiments}

\begin{table*}[t]
\centering
\footnotesize
\setlength{\tabcolsep}{2.5pt}
\renewcommand{\arraystretch}{0.9}

\resizebox{\textwidth}{!}{
\begin{tabular}{llcccccc}
\toprule

\multirow[c]{2}{*}{Category}
&
\multirow[c]{2}{*}{Method}
&
\multicolumn{2}{c}{STEW}
&
\multicolumn{2}{c}{EEGMAT}
&
\multicolumn{2}{c}{SELF}
\\

\cmidrule(lr){3-4}
\cmidrule(lr){5-6}
\cmidrule(lr){7-8}

& & ACC & F1 & ACC & F1 & ACC & F1 \\

\midrule

\multirow[c]{4}{*}{General EEG}
& EEGLearn
& $69.08\pm17.08$ & $63.41\pm23.04$
& $61.14\pm16.64$ & $56.23\pm20.20$
& $24.87\pm13.48$ & \underline{$22.59\pm12.04$}
\\

& EEGNet
& $73.00\pm14.04$ & $71.09\pm16.29$
& $67.66\pm13.18$ & $65.35\pm15.65$
& $20.42\pm2.93$ & $14.55\pm3.77$
\\

& EEG-Conformer
& \underline{$75.51\pm10.78$}
& \underline{$74.30\pm12.60$}
& $65.30\pm14.04$ & $61.68\pm16.99$
& $22.01\pm6.16$ & $18.16\pm7.14$
\\

& EEG-Deformer
& $74.30\pm13.26$ & $72.18\pm16.24$
& $63.36\pm9.63$ & $59.46\pm12.66$
& $19.85\pm8.35$ & $15.86\pm6.07$
\\

\midrule

\multirow[c]{4}{*}{Workload-Specific}
& LSCCN
& $68.28\pm11.95$ & $66.88\pm13.56$
& $57.25\pm9.36$ & $55.22\pm10.49$
& $24.64\pm4.56$ & $20.58\pm5.10$
\\

& MuLHiTA
& $71.22\pm14.00$ & $70.01\pm15.22$
& $62.57\pm9.90$ & $60.32\pm11.48$
& $20.97\pm2.98$ & $17.07\pm4.34$
\\

& EEGMeNet
& $73.21\pm14.10$ & $71.14\pm17.17$
& $66.46\pm12.83$ & $61.97\pm17.18$
& $21.85\pm8.21$ & $14.93\pm7.70$
\\

& BiConformer
& $73.69\pm15.53$ & $72.16\pm17.57$
& \underline{$70.14\pm13.49$}
& \underline{$66.97\pm17.06$}
& \underline{$25.87\pm6.97$}
& $19.20\pm5.36$
\\

\midrule

\multirow[c]{4}{*}{Graph-Based}
& DGCNN
& $74.83\pm13.78$ & $72.92\pm16.49$
& $66.39\pm13.28$ & $62.23\pm17.17$
& $24.62\pm7.23$ & $19.79\pm8.29$
\\

& RGNN
& $73.21\pm14.87$ & $70.97\pm17.98$
& $64.38\pm12.83$ & $60.32\pm15.98$
& $21.74\pm7.54$ & $14.37\pm7.68$
\\

& AdamGraph
& $73.45\pm15.10$ & $71.78\pm17.05$
& $65.09\pm14.02$ & $61.83\pm16.33$
& $23.08\pm4.75$ & $17.16\pm6.97$
\\

& DCGNN
& $71.00\pm13.02$ & $69.33\pm14.72$
& $63.37\pm13.26$ & $59.92\pm15.94$
& $24.81\pm6.21$ & $19.47\pm8.69$
\\

\midrule

Ours
& \textbf{ProtoGIB}
& \boldmath{$80.82\pm10.10$}
& \boldmath{$80.49\pm10.52$}
& \boldmath{$74.91\pm13.38$}
& \boldmath{$73.31\pm15.27$}
& \boldmath{$29.28\pm7.36$}
& \boldmath{$25.52\pm6.79$}
\\

\bottomrule
\end{tabular}
}

\caption{Cross-subject comparison on the three workload datasets.}
\label{tab:comparison-cs}
\end{table*}

\subsection{Experimental Settings}

This section introduces the datasets, comparison methods, evaluation protocols, and implementation details. The supplementary material provides further dataset statistics, baseline descriptions, dataset-specific evaluation procedures, and complete preprocessing and implementation configurations.

\subsubsection{Datasets}

We evaluate ProtoGIB-Workload on three EEG workload datasets. STEW~\cite{stew} includes 48 subjects, 14 electrodes, and 128-Hz recordings, with low and high workload defined by resting and multitasking conditions. EEGMAT~\cite{eegmat} includes 36 subjects, 19 electrodes, and 500-Hz recordings during rest and mental arithmetic. SELF is our self-collected dataset from 8 air traffic controllers performing five simulated air-traffic-control scenarios, containing 59-channel EEG recordings at 500 Hz and workload states ranging from rest to emergency handling. STEW and EEGMAT are widely used public benchmarks, while SELF extends the evaluation to a realistic professional task with finer-grained workload levels and greater recognition difficulty.

\subsubsection{Comparison Methods}

We compare ProtoGIB with twelve representative methods across three modeling paradigms. General EEG representation models include EEGLearn~\cite{eeglearn}, EEGNet~\cite{lawhern2018eegnet}, EEG-Conformer~\cite{song2023eegconformer}, and EEG-Deformer~\cite{ding2025eegdeformer}. Workload-oriented models include LSCCN~\cite{lsccn}, MuLHiTA~\cite{mulhita}, EEGMeNet~\cite{eegmenet}, and BiConformer~\cite{yan2025biconformer}. Graph-based EEG models include DGCNN~\cite{song2020dgcnn}, RGNN~\cite{zhong2022rgnn}, AdamGraph~\cite{philipchen2025adamgraph}, and DCGNN~\cite{xiao2025dcgnn}.

We focus on EEG-specific baselines because they are designed and validated for EEG representation learning or workload recognition. General-purpose architectures require task-specific redesigns of temporal encoders, graph construction, or signal preprocessing, making direct comparison less meaningful.

\subsubsection{Evaluation Protocols}

We evaluate within- and cross-subject settings. Within subjects, data are chronologically split into training, validation, and test sets at 70\%, 20\%, and 10\%. Across subjects, we use strict leave-one-subject-out validation, holding out one subject for testing and the rest for training and model selection. Cross-subject performance is primary because it measures generalization to unseen subjects. We report accuracy and Macro-F1, two widely used metrics in EEG workload recognition, using the latter as the primary metric. Within-subject results are averaged across subjects, while cross-subject results are averaged across LOSO folds.

\subsubsection{Preprocessing and Implementation Details}

Continuous EEG recordings are processed using a common pipeline across datasets. We divide the target signals into non-overlapping 1-second epochs. Each channel is independently normalized by z-score normalization. Dataset-specific bandpass filters and a 50 Hz notch filter are applied before segmentation. The original sampling rate of each dataset is retained.

All methods use the same processed data, splits, and evaluation protocols. ProtoGIB-Workload is optimized with Adam, with checkpoints and hyperparameters selected by validation Macro-F1. The search space, final settings, and analysis are reported in the Sensitivity Analysis section and supplementary material. The network architecture is fixed across datasets, while temporal input length follows the native sampling rate. Baselines use released code or reported configurations when available. Model selection relies only on validation data, without test access. The random seed is set to 2026, and the hardware and software environment is detailed in the supplementary material.

\subsubsection{Comparison Experiments}

We conduct a systematic comparison between ProtoGIB and twelve representative methods on STEW, EEGMAT, and SELF under both within-subject and cross-subject settings. Performance is evaluated using ACC and Macro-F1. Values are reported as percentages in the form of mean $\pm$ standard deviation. Tables~\ref{tab:comparison-ws} and~\ref{tab:comparison-cs} present the within-subject and cross-subject results, respectively. The best result in each setting is highlighted in bold, while the second-best result is underlined. The results lead to three main observations.

\noindent\textbf{Observation 1. ProtoGIB delivers consistently strong performance across datasets and evaluation protocols.} ProtoGIB ranks first in both ACC and Macro-F1 on all three datasets under within-subject and cross-subject settings. Under within-subject evaluation, it improves the strongest competing Macro-F1 score by 1.50, 2.98, and 3.71 percentage points on STEW, EEGMAT, and SELF, respectively, with an average gain of 2.73 points. Under cross-subject evaluation, the gains increase to 6.19, 6.34, and 2.93 points, averaging 5.15 points. This consistency across binary and multi-class tasks, electrode systems, and sampling rates suggests that the framework can generalize across different acquisition settings and workload formulations.

\noindent\textbf{Observation 2. The advantage of ProtoGIB becomes more pronounced when generalizing to unseen subjects.} The average Macro-F1 improvement under cross-subject evaluation is nearly twice that under within-subject evaluation. On STEW and EEGMAT, ProtoGIB exceeds the strongest baselines by 6.19 and 6.34 points under LOSO, compared with 1.50 and 2.98 points within subjects. Within-subject models may exploit recurring individual patterns that do not transfer, whereas LOSO exposes subject-dependent structural shortcuts. The larger gains under LOSO are consistent with the motivation of ProtoGIB to reduce subject-dependent connectivity variation. 

\begin{figure*}[!t]
\centering
\begin{subfigure}[t]{0.32\textwidth}
    \centering
    \includegraphics[width=\linewidth]{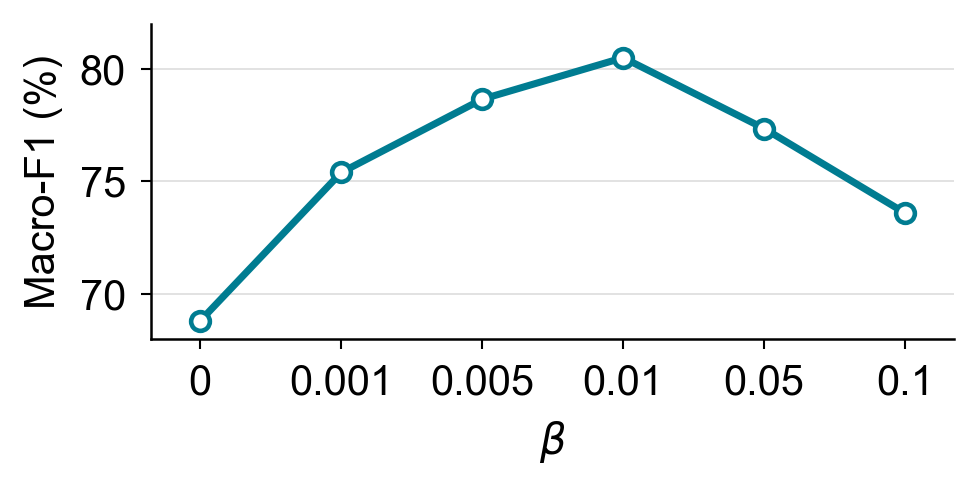}
    \caption{Information bottleneck weight $\beta$.}
    \label{fig:sensitivity-beta}
\end{subfigure}
\hfill
\begin{subfigure}[t]{0.32\textwidth}
    \centering
    \includegraphics[width=\linewidth]{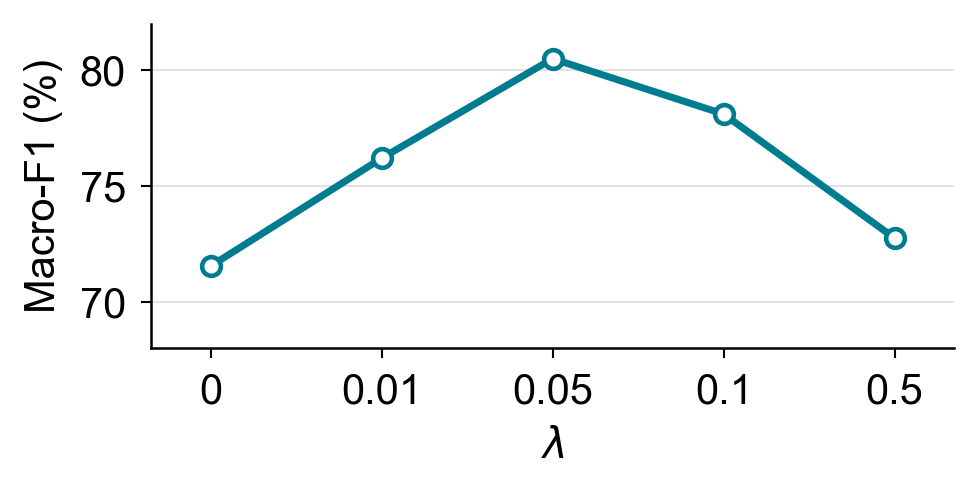}
    \caption{Topology stabilization weight $\lambda$.}
    \label{fig:sensitivity-lambda}
\end{subfigure}
\hfill
\begin{subfigure}[t]{0.32\textwidth}
    \centering
    \includegraphics[width=\linewidth]{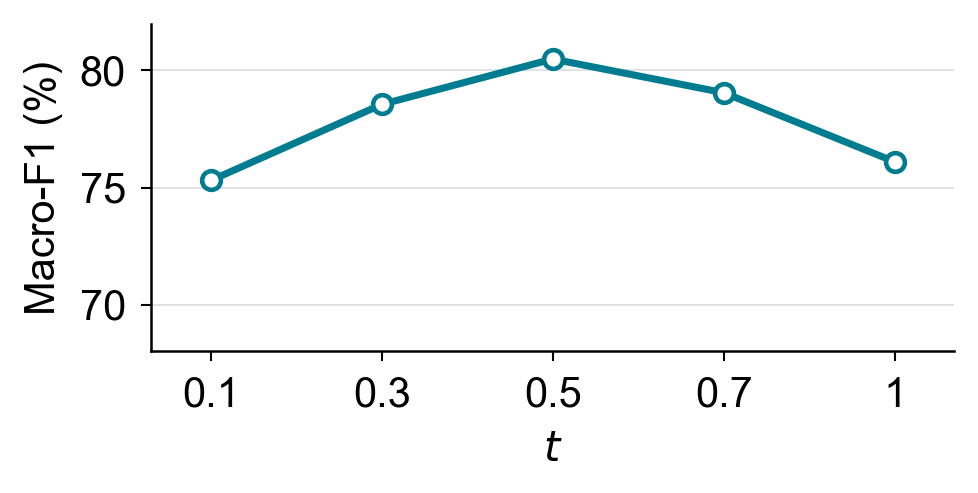}
    \caption{Gumbel-Sigmoid temperature $t$.}
    \label{fig:sensitivity-temperature}
\end{subfigure}
\caption{Sensitivity of ProtoGIB to the main hyperparameters on STEW under cross-subject evaluation.}
\label{fig:sensitivity}
\end{figure*}

\noindent\textbf{Observation 3. Explicit graph modeling alone is insufficient to resolve subject-dependent topology variation.} The graph-based baselines directly model inter-channel relationships, yet all underperform ProtoGIB in cross-subject evaluation. Compared with DGCNN, the strongest graph baseline by cross-subject Macro-F1, ProtoGIB gains 7.57 points on STEW, 11.08 on EEGMAT, and 5.73 on SELF. This suggests that explicitly modeling inter-channel relationships alone may be insufficient for cross-subject generalization, and that controlling subject-dependent topology variation can provide additional benefits. SELF is especially challenging: with five classes but only eight subjects, all methods degrade substantially under LOSO. ProtoGIB still achieves the best Macro-F1 of 25.52\%, though the low absolute score indicates that fine-grained cross-subject recognition remains difficult with limited subject diversity.

To examine whether the observed improvements are consistent across subjects, we conduct paired Wilcoxon signed-rank tests between ProtoGIB and every baseline using subject-level ACC and Macro-F1 scores. The analysis covers both within-subject and cross-subject evaluation. More than 80\% of the comparisons yield $p$-values below 0.05, indicating that the gains of ProtoGIB are consistent across subjects rather than being driven by a small number of cases. Significance results are provided in the supplementary material.

\subsubsection{Ablation Study}

We conduct ablation experiments to evaluate the individual contributions of the Stochastic Graph Information Bottleneck (SGIB) and the Class-Conditional Topology Stabilizer (CTS). Specifically, we remove SGIB or CTS individually to obtain the w/o SGIB and w/o CTS variants, respectively, and remove both components to construct the Base model. All other experimental settings remain unchanged. Table~\ref{tab:ablation-cs-main} reports the cross-subject ACC and Macro-F1 results on the three datasets in terms of mean $\pm$ standard deviation. The complete within-subject ablation results are provided in the supplementary material.

\begin{table}[!t]
\centering
\footnotesize
\setlength{\tabcolsep}{3.0pt}
\renewcommand{\arraystretch}{0.9}
\resizebox{\columnwidth}{!}{
\begin{tabular}{llccc}
\toprule
Model & Metric & STEW & EEGMAT & SELF \\
\midrule

\multirow{2}{*}{\textbf{Ours}}
& ACC
& {\boldmath$80.82\pm10.10$}
& {\boldmath$74.91\pm13.38$}
& {\boldmath$29.28\pm7.36$}
\\
& F1
& {\boldmath$80.49\pm10.52$}
& {\boldmath$73.31\pm15.27$}
& {\boldmath$25.52\pm6.79$}
\\

\midrule

\multirow{2}{*}{w/o SGIB}
& ACC
& $71.02\pm14.08$
& $65.98\pm14.89$
& $29.04\pm7.16$
\\
& F1
& $68.82\pm16.83$
& $65.98\pm18.44$
& $24.48\pm6.98$
\\

\midrule

\multirow{2}{*}{w/o CTS}
& ACC
& $73.06\pm16.06$
& $70.21\pm14.83$
& $27.18\pm5.31$
\\
& F1
& $71.55\pm17.84$
& $66.50\pm18.86$
& $22.76\pm5.80$
\\

\midrule

\multirow{2}{*}{Base}
& ACC
& $68.09\pm17.35$
& $63.15\pm13.66$
& $23.47\pm6.41$
\\
& F1
& $65.11\pm20.52$
& $58.70\pm16.80$
& $16.64\pm8.00$
\\

\bottomrule
\end{tabular}
}
\caption{Cross-subject ablation results on the three datasets.}
\label{tab:ablation-cs-main}
\end{table}

The full model performs best across all datasets and metrics. Removing either component degrades performance, while removing both causes the largest reduction, confirming their complementary roles. The information bottleneck has a larger effect on STEW and EEGMAT, reducing Macro-F1 by 11.67 and 7.33 points when removed. CTS contributes more on SELF, where its removal lowers Macro-F1 by 2.76 points, compared with 1.04 points for the information bottleneck. Thus, their relative importance varies across datasets, but their combination consistently yields the best cross-subject performance.

\subsubsection{Sensitivity Analysis}

We evaluate the sensitivity of ProtoGIB to the information bottleneck weight $\beta$, topology stabilization weight $\lambda$, and Gumbel-Sigmoid temperature $t$. One parameter is varied at a time while the others remain fixed. We test $\beta\in\{0,0.001,0.005,0.01,0.05,0.1\}$, $\lambda\in\{0,0.01,0.05,0.1,0.5\}$, and $t\in\{0.1,0.3,0.5,0.7,1.0\}$ on STEW under cross-subject evaluation. The results are shown in Fig.~\ref{fig:sensitivity}.

The best performance occurs at $\beta=0.01$, $\lambda=0.05$, and $t=0.5$. The small $\beta$ suggests that mild compression suppresses redundant connectivity while preserving workload-discriminative structures. The moderate $\lambda$ improves cross-subject topology consistency, whereas stronger regularization may suppress meaningful class-dependent variation. The intermediate $t=0.5$ balances discrete edge selection with stable gradient optimization.

\begin{figure}[!t]
\centering
\includegraphics[width=0.8\columnwidth]{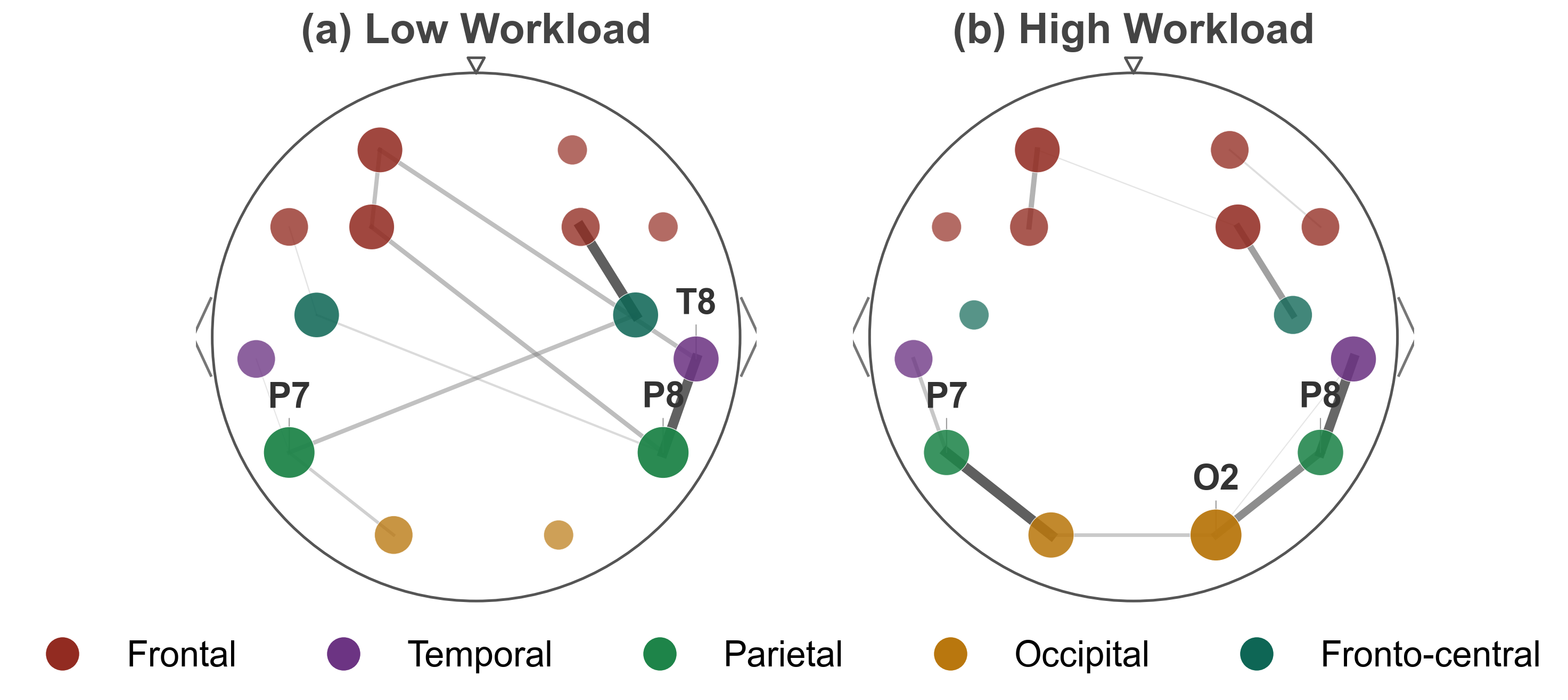}
\caption{Topology prototypes and channel relevance.}
\label{fig:connectivity-vis}
\end{figure}

\subsubsection{Prototype Topology and Channel Relevance}

We analyze workload-specific connectivity learned by ProtoGIB on STEW under cross-subject evaluation. Figure~\ref{fig:connectivity-vis} shows the class-conditional topology prototypes, where edge strength denotes normalized class-averaged retention weights and node size denotes channel weighted degree. Low workload exhibits distributed frontal, fronto-central, and parietal connectivity, whereas high workload shows stronger posterior connectivity involving P7, P8, and O2 while retaining frontal interactions. These patterns may reflect greater posterior visual-attentional involvement during multitasking and provide qualitative evidence that ProtoGIB captures class-discriminative, interpretable connectivity.

        \section{Conclusion}
    
In this paper, we propose ProtoGIB-Workload, a graph bottleneck framework for subject-independent EEG workload recognition. To mitigate inter-subject heterogeneity, it integrates a Stochastic Graph Information Bottleneck (SGIB) to extract compact, task-relevant subgraphs, and a Class-Conditional Topology Stabilizer (CTS) to align subject-level edge profiles toward stable, workload-specific prototypes. Theoretical analyses establish its effectiveness in bounding source-subject information. Extensive evaluations across three datasets, including an in-house air traffic controller dataset, demonstrate that ProtoGIB-Workload achieves state-of-the-art cross-subject generalization. Future work will extend this framework to dynamic temporal graphs for continuous monitoring in safety-critical BCI applications.

    \bibliography{aaai2027}

    \end{document}